\documentclass[12pt,preprint]{aastex}

\shorttitle{The Spatially Resolved H$\alpha$-Emitting\\
		Wind Structure of P Cygni}
\shortauthors{Balan et al.}

\begin{document}

\title{The Spatially Resolved H$\alpha$-Emitting Wind Structure of P Cygni}

\author{Aurelian Balan\altaffilmark{1,2}, C. Tycner\altaffilmark{1},
  R. T. Zavala\altaffilmark{3}, J. A. Benson\altaffilmark{3},
  D. J. Hutter\altaffilmark{3}, M. Templeton\altaffilmark{4}}

\altaffiltext{1}{Department of Physics, Central Michigan University,
  Mount Pleasant, MI 48859; abalan@delta.edu, c.tycner@cmich.edu.}
\altaffiltext{2}{Current address: Delta College, Science Division,
  University Center, MI, 48710} 
\altaffiltext{3}{US Naval Observatory, Flagstaff Station, 10391 W.
  Naval Observatory Rd., Flagstaff, AZ 86001; bzavala@nofs.navy.mil,
  jbenson@nofs.navy.mil, djh@nofs.navy.mil} 
\altaffiltext{4}{American Association of Variable Star Observers, 49
  Bay State Road, Cambridge, MA 02138; matthewt@aavso.org}

\begin{abstract}
High spatial resolution observations of the H$\alpha$-emitting wind
structure associated with the Luminous Blue Variable star P~Cygni were
obtained with the Navy Prototype Optical Interferometer (NPOI). These
observations represent the most comprehensive interferometric data set
on P~Cyg to date. We demonstrate how the apparent size of the
H$\alpha$-emitting region of the wind structure of P~Cyg compares
between the 2005, 2007 and 2008 observing seasons and how this relates
to the H$\alpha$ line spectroscopy. Using the data sets from 2005,
2007 and 2008 observing seasons, we fit a circularly symmetric
Gaussian model to the interferometric signature from the
H$\alpha$-emitting wind structure of P~Cyg.  Based on our results we
conclude that the radial extent of the H$\alpha$-emitting wind
structure around P~Cyg is stable at the 10\% level. We also show how
the radial distribution of the H$\alpha$ flux from the wind structure
deviates from a Gaussian shape, whereas a two-component Gaussian model
is sufficient to fully describe the H$\alpha$-emitting region around
P~Cyg.
\end{abstract}


\section{Introduction}

Luminous Blue Variable (LBV) star P~Cyg (HD~193237, B2pe) is an
unusual star with a unique stellar wind structure. LBVs are evolved,
very luminous, massive supergiant stars that show some type of
instability. As the likely progenitors of Wolf-Rayet
stars~\citep{crowther07} LBV's may provide insight for the ultimate
end-stage of these stars, the core-collapse supernovae.  P~Cyg was
discovered in 1600 after a violent mass-loss event caused it to
quickly brighten to a third magnitude star \citep{deg69}. The violent
mass-loss events that spur these eruptions are not fully understood,
but a number of competing models including single and binary star
scenarios attempt to explain these eruptions \citep{hum94}. With a
mass-loss between 2$\times$10$^{-5}$ and 4$\times$10$^{-4}$
$M_{\odot}$~$\rm yr^{-1}$ \citep{vanb78,abbott80,lei87}, P~Cyg
displays characteristics that are suggestive of energetic mass
outflows. It is for this reason that P~Cyg has a unique stellar wind
structure with possible high density regions in motion within the wind
structure. Despite numerous studies published in the literature that
discuss P~Cyg or other LBVs, there is still need for observations that
can spatially resolve the circumstellar region around P~Cyg.

The spectrum of P~Cyg reveals emission line profiles with their
archetypical shapes, which are formed by the circumstellar matter that
has been shed from the central star. These profiles are commonly
identified by an emission component to the red side of an absorption
line. The origin of this Doppler shift can be explained with a
spherically symmetric outflowing wind in which the velocity increases
with radial distance~\citep[see, e.g.,][\S~2.2]{lam99}.  The
absorption component corresponds to the continuum light absorbed by
the wind structure directly between the observer and the central star.
The emission component on the other hand originates in the spherically
symmetric halo around the central star.  Because the same
characteristics apply to the H$\alpha$ emission line, which is one of
the strongest emission lines in the spectrum of P~Cyg, this makes the
H$\alpha$ emission line an excellent probe of the wind region around
P~Cyg.

P~Cyg currently has an estimated mass of
30~$\pm$~10~$M_{\odot}$~\citep{lam83, lam85}. At the start of its
lifetime P~Cyg possibly had an initial mass of 50 $M_{\odot}$
\citep{deg92}, but has shed much of its mass because of its violent
history of mass loss events. The effective temperature and the stellar
diameter of P~Cyg are estimated to be $\rm T_{\rm
  eff}$~=~19,300~$\pm$~2000~K and 76~$\pm$~14~$R_{\odot}$,
respectively \citep{lam83}. However, the H$\alpha$-emitting region
extends radially much farther than the central star allowing us to
resolve this region using long-baseline interferometric techniques,
even though the distance is estimated to be in the range of
$1.7~\pm~0.1$~kpc to $1.8\pm~0.1$~kpc \citep{lam83, naj97}.

The circumstellar region of P~Cyg has been spatially resolved in the
past using radio and optical interferometry, as well as direct imaging
with adaptive optics~(AO).  Radio interferometric observations detect
the nebula around P~Cyg at the angular scales from $\sim$50~mas to
almost an arcminute~\citep{skin97,skin98}, whereas optical
long-baseline interferometry is sensitive to structures in the nebula
of P~Cyg that are more than an order of magnitude smaller.  Therefore,
optical interferometry provides a unique window of opportunity to
resolve the inner portion of the wind structure.

Most relevant to our study are the results that probed the
H$\alpha$-emitting region of P~Cyg.  For example, based on
near-diffraction limited observations obtained using AO system on a
1.52~m telescope, \citet{che00} not only resolved the outer
H$\alpha$-emitting region of the extended envelope, but detected
signatures of clumping.  Although, the angular scales sampled with a
1.52~m telescope were quite large, in the region of ~600~mas, the
angular resolution was limited to 50~mas.  On much smaller angular
scales, \citet{vak97} using a single 17.7~m baseline of the GI2T
interferometer have resolved the inner H$\alpha$-emitting envelope of
P~Cyg yielding an angular size of 5.52$\;\pm\;$0.47 mas and reported a
marginal detection at the HeI~6678 line.  The spatial scales probed by
our study are similar to those sampled by \citet{vak97}, however, the
interferometric observations at the H$\alpha$ line obtained in our
study were acquired using 11 unique interferometric baselines ranging
in length from 15.8 to 79.3~m, thus resulting in much higher angular
resolution measurements.

\section{Observations}

\subsection{Interferometry} \label{inter}

High angular resolution long-baseline interferometric observations
were obtained on P~Cyg over thirty-seven nights in 2005, 2007 and 2008
using the Navy Prototype Optical Interferometer (NPOI) located near
Flagstaff, Arizona.  The NPOI is a long-baseline optical
interferometer which consists of six 50~cm siderostat mirrors that
send 12.5~cm diameter light beams down the optical feed system
consisting of 15~cm diameter pipes to an optics lab where the
interference~(i.e., a fringe) between the beams is
recorded~\citep{arm98}. Observations covering a wide spectral band can
be taken over multiple baselines (currently ranging from 19 to 80~m in
length), and are geometrically compensated and modulated using vacuum 
delay-lines to guarantee the entire band is being observed simultaneously.

The observing process involves measuring the fringe contrast over
15~spectral channels every 2~ms. Each 2~ms frame is then used to
obtain a squared visibility~($V^2$) measure~(i.e., a V$^2$ value
measures the fringe contrast). Afterward, these values are
incoherently averaged over 1~s intervals to create 1~s data points
\citep{hum03}. The final observations consist of 30~s of integration
(also known as a "scan") during which the squared visibilities are
measured simultaneously at three baselines per output from the
beam-combiner (with two outputs being used simultaneously) and 15
spectral channels covering the wavelengths between 560 and 870~nm. The
typical observational sequence consists of a pair of coherent and
incoherent (off the fringe) scans on a target followed by a pair of
scans on a calibrator. The incoherent scans are used to estimate the
additive bias terms affecting the squared visibility measures. The
process is repeated during a night for as long as the fringe packet
can be detected on the target star. Typically, the observations stop
when the optical path difference between the different siderostats
cannot be compensated anymore.

The H$\alpha$-emitting sources form a specific class of targets for
NPOI since the entire emission line ends up only in one out of the 15
spectral channels. This characteristic allows one to calibrate the
$V^2$ values from the channel containing the H$\alpha$ emission using
the $V^2$ values from the neighboring continuum
channels~\citep{Tycner03}.  Although during this calibration process
one needs to assume that the diameter of the central star at the
continuum wavelengths is known, the final results based on the
H$\alpha$ emission are only weakly dependent on the adopted value for
the angular diameter of the central star (we explore this in more
detail in \S~\ref{central}). For the purpose of the calibration we
adopted a uniform disk diameter of 0.2~mas to represent the central
star at the continuum wavelengths. We obtain this diameter based on a
distance estimate of $1.8\pm~0.1$~kpc and a central star radius of
76~$\pm$~14~$R_{\odot}$ \citep{lam83}.

\begin{figure}
\plotone{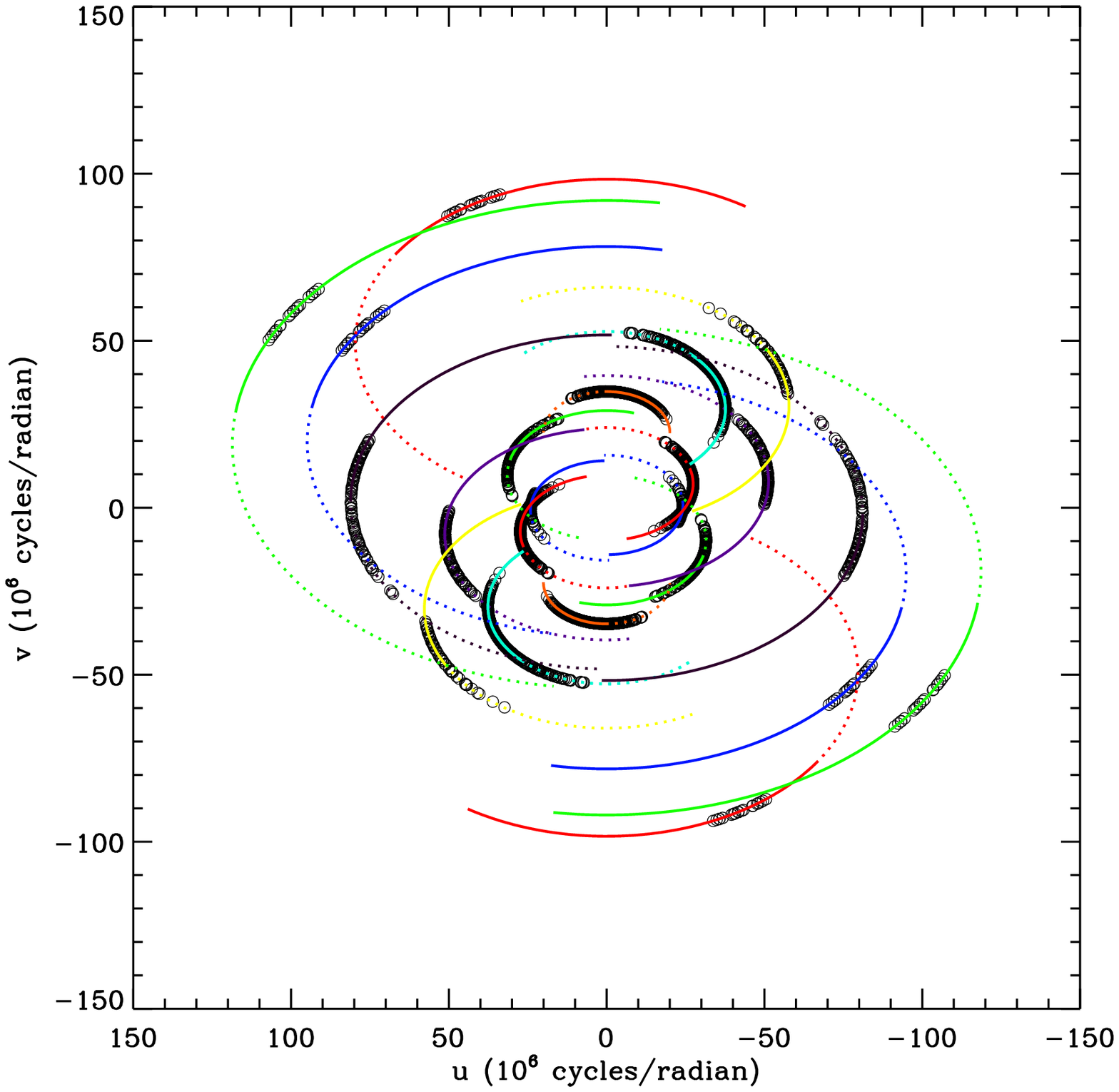}
\caption{Sampling of the $(u,v)$-plane in the H$\alpha$ channel for P
  Cyg on multiple baselines ranging in length from 15.8 to 79.3 m
  obtained on thirty-seven nights of observation in 2005, 2007 and
  2008 ({\it open circles}).  Possible coverage at each baseline is
  also shown from 6~hr east of meridian~({\it dotted line}) to 6~hr
  west of meridian~({\it solid line}). \label{all_uv_sampling}}
\end{figure}

Applying the procedures described above to the entire observational
data set we acquired for P~Cyg, yielded a total of 1,534~H$\alpha$
squared visibilities on 11 unique baselines~(out of a maximum of 15
unique baselines for a 6-station configuration).  Table~\ref{log}
lists the number of unique H$\alpha$ $V^2$ measures we obtained for
each of the nights in our observing campaign.  Although the
observational setup for each of the three observing seasons was
slightly different, and therefore a range of different baselines were
sampled each year, in total we have obtained observations at baselines
ranging from the shortest~(15.8~m) to the longest~(79.3~m) possible at
the time of our observations.  More specifically, the 2005 and 2008
seasons contained baselines between 18.9 and 53.2~m, and the 2007
season contained baselines from 15.8 to 79.3~m in length.
Figure~\ref{all_uv_sampling} shows the sampling of the $(u,v)$-plane
in the H$\alpha$ channel for P~Cyg based on all the available nights
in 2005, 2007 and 2008. The ``arcs'' seen in the plot are a result of
the changing projection of the baseline vector on the sky during the
night due to Earth's rotation. Three seasons of squared visibility
data sampled across the $(u,v)$-plane are shown in
Figure~\ref{fig:all_v2} plotted as a function of radial spatial
frequency ($\sqrt{u^{2}+v^{2}}$).  The calibrated H$\alpha$ $V^2$
measures are listed in Table~\ref{tab:v2data}.

In addition to the $V^2$ measures we have also obtained closure phase
quantities (sum of complex phases on three baselines forming a
triangle) on four unique baseline triangles.  By fitting a quadratic
function to the closure phases from the continuum channels~(at each
scan and closure phase configuration separately), and then removing
this quadratic trend from all channels, including the H$\alpha$
channel, we were able to obtain the differential closure phases for
the H$\alpha$-emitting region.  Because we expect the central star to
be mostly unresolved at the continuum channels and if we assume that
the continuum source is point-symmetric, then we expect the closure
phases to cluster around a zero value.  This allows us to inspect the
H$\alpha$ closure phases with respect to the mean zero closure phase.
In order to reduce the random noise present in data associated with a
single scan we calculated the weighted mean closure phases for each
season and each unique configuration.  In the 2007 season one of the
baselines (AE--AN) was sampled at two independent outputs from the
beam combiner and this allowed us to calculate two closure phase
quantities containing this baseline either from one output beam or the
other.  Because the observations for the other baselines were
identical, we decided not to average these quantities, and instead
treat them separately.  Figure~\ref{fig:closure_phases} shows the
weighted mean closure phases for all the configurations with the
quadratic trend based on the continuum channels already removed.

\subsection{Spectroscopy} \label{spec}

\begin{figure}
\plotone{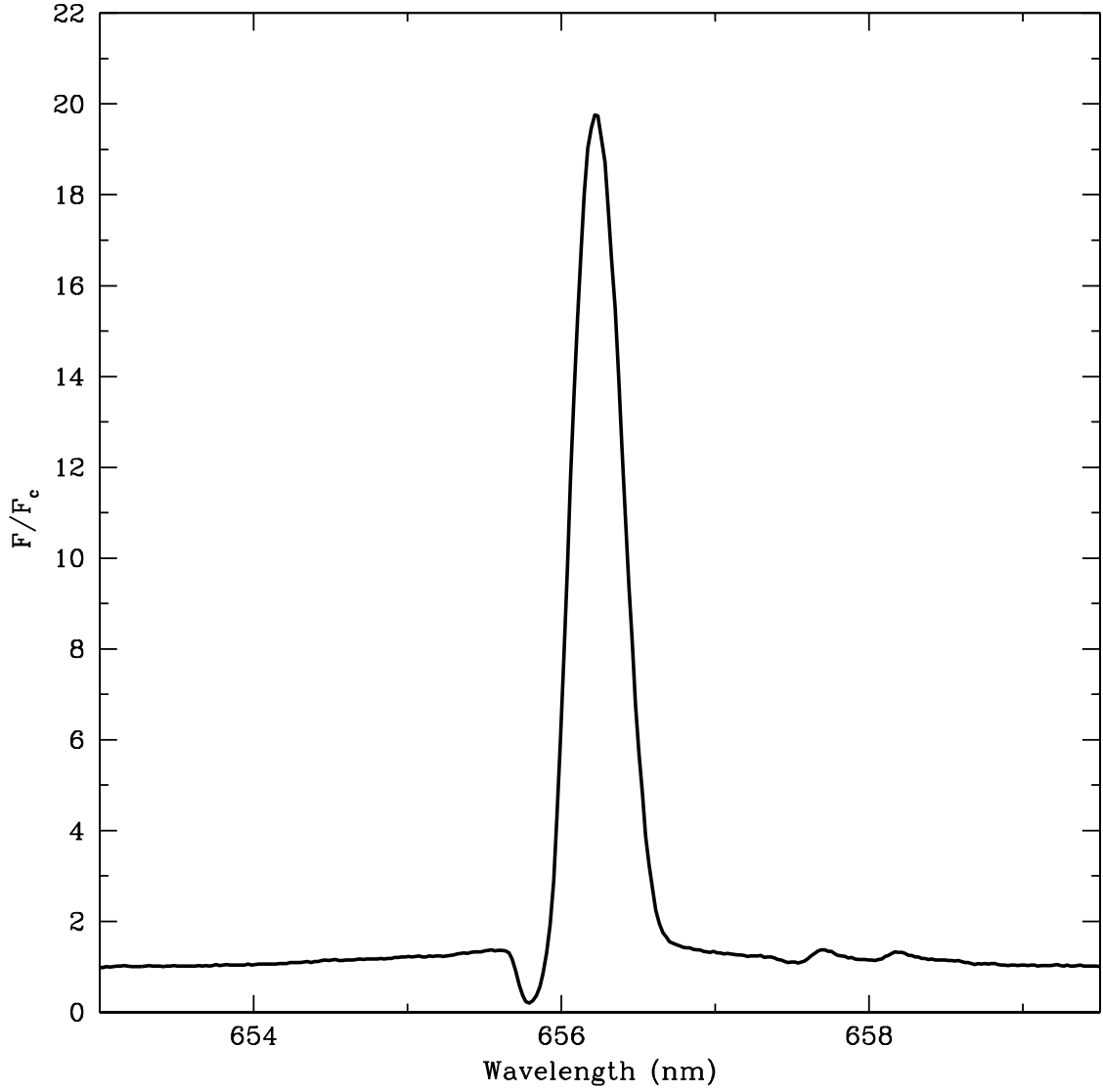}
\caption{A representative H$\alpha$ line profile of P Cyg obtained on 2007
  June 22. The equivalent width of the emission line is $-80.0$~\AA.}
\label{fig:Pcyg-spectra}
\end{figure}

\begin{figure}
\plotone{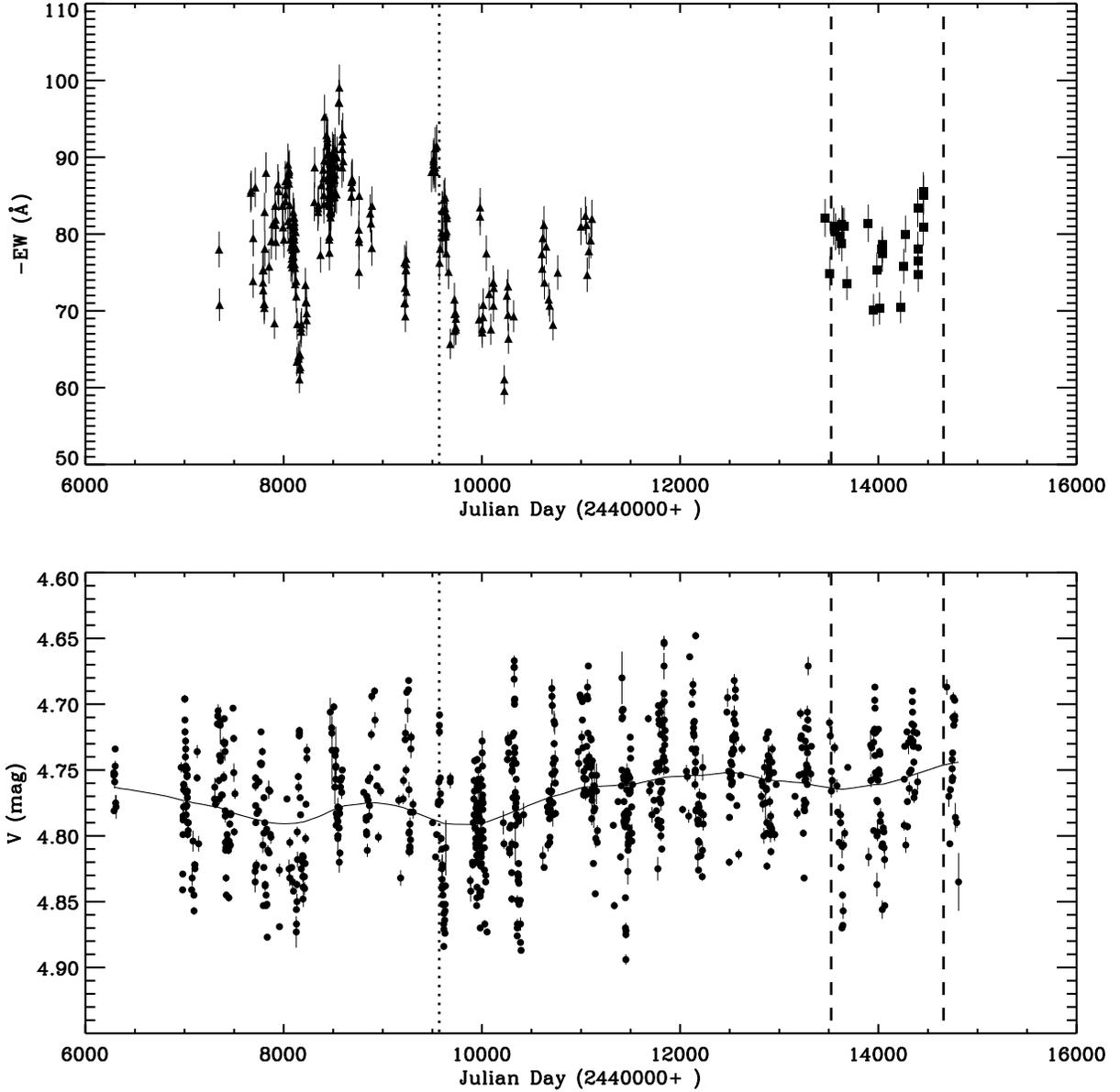}
\caption{{\it Upper panel}: P~Cyg H$\alpha$ equivalent widths from
  1988 Jul to 2007 Dec with data from \citet{mar01a}~({\it triangles})
  and this study~({\it squares}) shown.  The vertical lines mark the
  periods when interferometric observations were obtained at the NPOI
  ({\it dashed lines}), and the single night of interferometric
  observation of P~Cyg reported by \citet{vak97}~({\it dotted line}).
  {\it Lower panel}: The AAVSO Johnson V-band photoelectric light
  curve of P~Cyg for the period 2005--2009, obtained based on
  measurements relative to the comparison star HD 188892 ($V=4.936$,
  $B-V=-0.086$).  Only data where the measured check star magnitude
  (HD 193369; $V=5.573$) lies within $\pm 0.04$ mag of the known value
  are plotted.  Typical photometric errors per point are 5--10
  milli-magnitudes.  The smoothed moving average (with a 1000~d
  window) is also shown ({\it solid line}) to guide the eye for
  possible long-term variability.
  \label{fig:EW_vs_V}}
\end{figure}

The H$\alpha$ emission line, as detected by the NPOI, falls on a
spectral channel that has a spectral width of 15~nm.  Therefore, the
interferometric data obtained on P~Cyg does not provide sufficient
spectroscopic information to allow us to determine the strength or the
shape of the H$\alpha$ emission line.  For this reason, we obtained
complementary spectroscopy of P~Cyg using the Lowell Observatory's
John S. Hall 1.1~m telescope located at the same observing site as the
NPOI.  The telescope was outfitted with the Solar-Stellar Spectrograph
(SSS), which is a fiber-fed echelle spectrograph~\citep{Hall95}.
Spectroscopic data on P~Cyg were obtained during the period from April
2005 to December 2007, with the individual nights listed in
Table~\ref{specvartab}. The SSS instrument produces spectra in the
H$\alpha$ region with a resolving power of $\sim$10,000.  A typical
H$\alpha$ line profile normalized with respect to the continuum level
of P~Cyg is shown in Figure~\ref{fig:Pcyg-spectra}.  The equivalent
width~(EW) measurements based on the individual spectra are also
plotted in the upper panel of Figure~\ref{fig:EW_vs_V}.

\section{Analysis}

\subsection{Interferometric Results} \label{observation}

\begin{figure}
\plotone{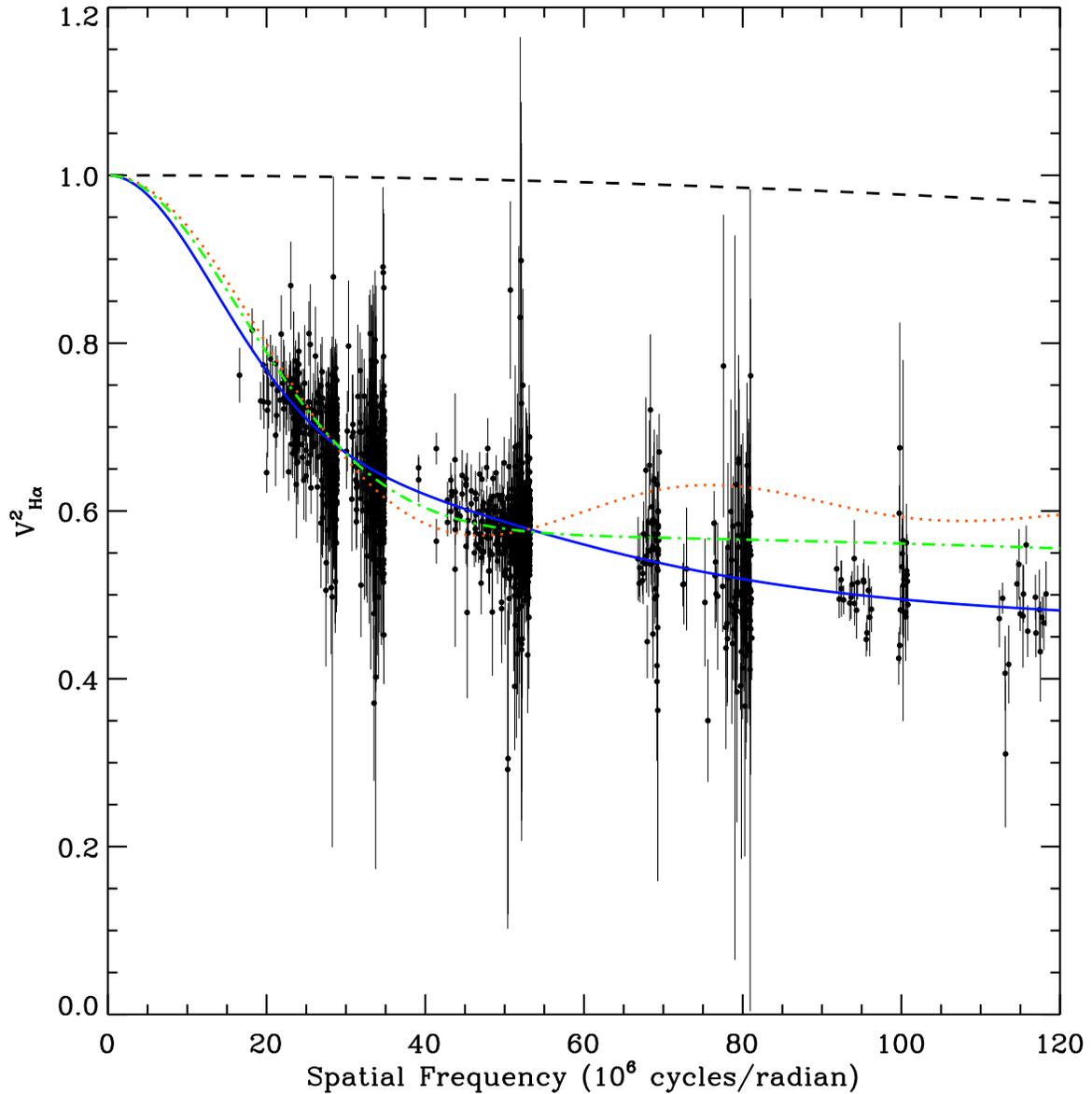}
\caption{The calibrated squared visibilities from the H$\alpha$
  channel of P~Cyg obtained on 11 unique baselines from 37 nights of
  observation in 2005, 2007 and 2008.  A signature of a central star
  represented by a UD model with angular diameter of 0.2~mas is
  shown~({\it dashed-line}) along with three models representing the
  H$\alpha$-emitting envelope.  The envelope component is modeled with
  a uniform disk~({\it dotted-line}), a single component
  Gaussian~({\it dash-dotted-line}), and a two-component Gaussian
  model~({\it solid-line}).
\label{fig:all_v2}}
\end{figure}

There are two signatures in the squared visibilities from the spectral
channel containing the H$\alpha$ emission line shown in Figure
\ref{fig:all_v2}. One is from the central star and the other is from the
H$\alpha$-emitting circumstellar region. To model this data, we use
the same two component model as was used by~\citet{Tycner06} of the
form:
\begin{equation} 
\label{eq:model1}
V^{2}_{\rm model}=[c_{\rm p}V_{\rm p}+(1-c_{\rm p})V_{\rm env}]^{2},
\end{equation}
where $V_{\rm p}$ is the visibility function representing the
photosphere of the central star, $V_{\rm env}$ is the visibility
function representing the circumstellar envelope, and $c_{\rm p}$ is
the fractional contribution from the stellar photosphere to the total
flux in the H$\alpha$ channel.  

Based on the results from differential closure phases~(recall
\S~\ref{inter}), which showed very weak (if any) variations across the
H$\alpha$ channel, we expect the H$\alpha$-emitting structure to be
sufficiently well described (at the level of angular resolution
provided by our range of baselines) by a model that is both point
symmetric and concentric with the central star.  This means that
$V_{\rm p}$ and $V_{\rm env}$ in equation~\ref{eq:model1} can be
represented by real functions (i.e., the Fourier transform of real and
even function results in a real function).  To model the stellar
photosphere we use a circularly symmetric uniform disk brightness
distribution. The normalized visibility amplitude for a uniform disk
can be written as:
\begin{equation} 
\label{eq:model2}
V_{\rm p}(u,v)={2\frac{J_{1}(\pi a s)}{\pi a s}},
\end{equation}
where $J_{1}$ denotes the Bessel function of the first kind and first
order, $a$ is the angular diameter, and $s$ is the radial spatial
frequency~(i.e., $s = \sqrt{u^{2}+v^{2}}$).  Because the
interferometric observations of P~Cyg were obtained at sufficiently
high spatial frequencies so that the H$\alpha$-emitting region was
fully resolved, we can compare different models of brightness
distribution for the envelope component.  We choose to model our data
with a uniform disk (UD) and a Gaussian distribution (GD), where in
the former $V_{\rm env}$ has the same functional form as
equation~\ref{eq:model2} except that $a$ is replaced with $\theta_{\rm
  UD}$, and in the latter case
\begin{equation} 
\label{eq:model4}
V_{\rm env}^{\rm GD}= \exp\left[{-\frac{(\pi \theta_{\rm GD} s)^2}{4
      \ln 2}}\right],\\
\end{equation}
where $\theta_{\rm GD}$ corresponds to the full-width at half-maximum
(FWHM) of the Gaussian model.

\begin{figure}
\plotone{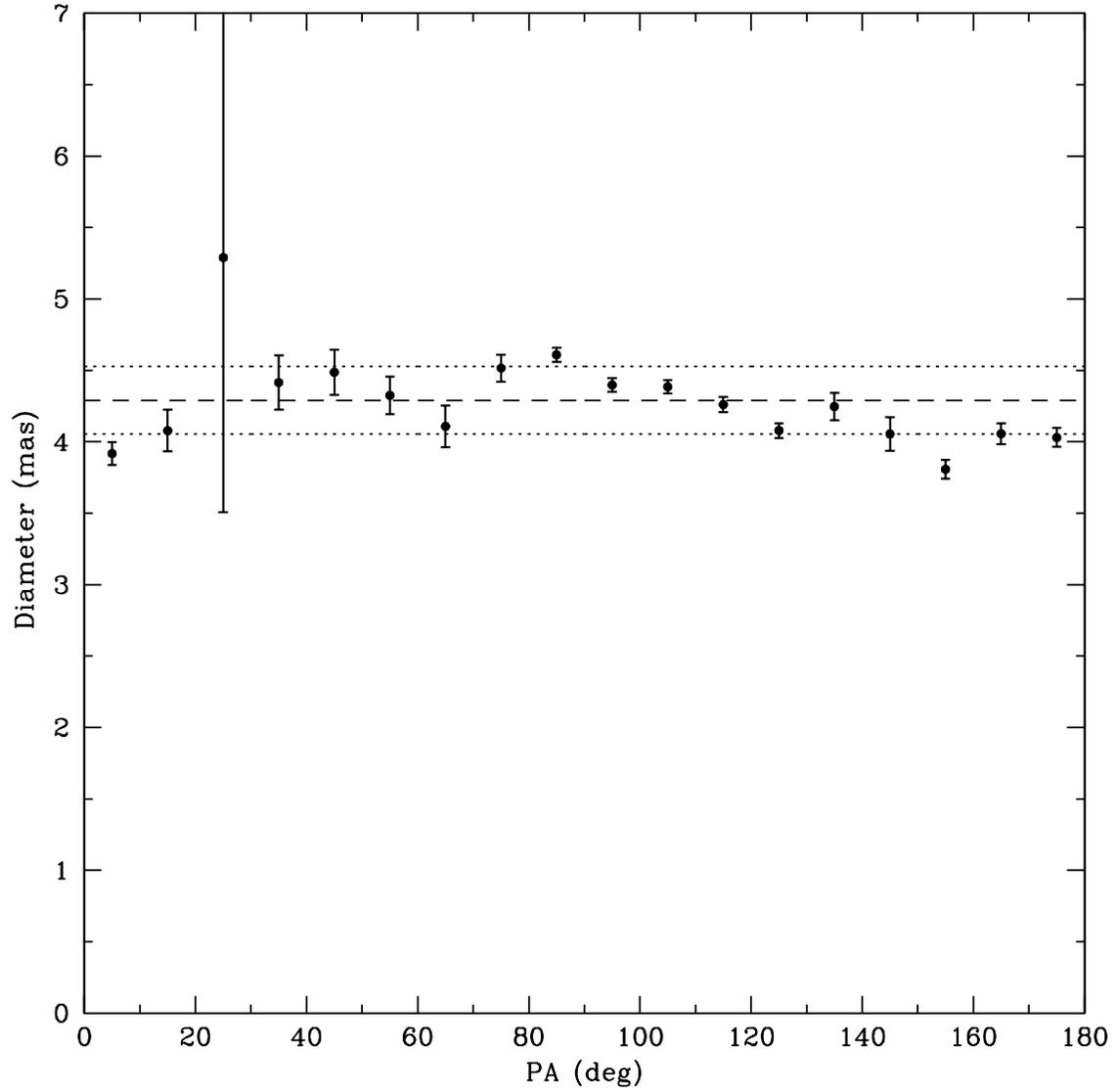}
\caption{Diameter of a Gaussian model fitted to subsets of squared
  visibility data divided based on ranges of PA on the sky.  Each
  point represents a 10$^{\circ}$ range in PA.  The mean diameter of
  4.29$\pm$0.03~mas based on a model fit to all data ({\it dashed
    line}) and 5.5\% variation around that value ({\it dotted lines})
  are also shown.  }
\label{fig:size_vs_PA}
\end{figure}

The envelope component~($V_{\rm env}$) in equation~\ref{eq:model1},
represented by either a uniform disk or a Gaussian radial
distribution, was assumed to be circularly symmetric.  This choice was
based on a test performed on the observational data set being divided
into 18 subsets based on the positional angles covered by the
observations.  Figure~\ref{fig:size_vs_PA} shows the variation of the
fitted diameter of a Gaussian model as a function of PA.  Although a
hint of variation is present in the data, more than 68\% (based on
$\pm 1 \sigma$ expectation) of points fall within 5.5\% of the mean
value and thus we conclude that some of that variation might be solely
due to the fact that the observations were acquired over many
observing seasons and intrinsic variability at the 5\% level cannot be
ruled out.  Similar test performed on much smaller data set from only
one season does not reveal any variation beyond those expected based
on the uncertainties associated with the fitted model parameters.
Therefore, for the purpose of describing the geometrical structure of
the H$\alpha$-emitting region of P~Cyg, we decided to concentrate only
on circularly symmetric models fitted to the $V^2$ data from the
individual observing seasons, as well as to the entire data set.

Table~\ref{modelpar} summarizes both UD and GD model results along
with their corresponding reduced $\chi^2$ values ($\chi^2_{\nu}$),
which were used to assess the goodness-of-fit of each model.  Based on
the $\chi^2_{\nu}$ values obtained from model fits to the entire data
set, it is clearly evident that the Gaussian model (with $\chi^2_{\nu}
= 1.9$) produced a significantly better fit to the data than a
circularly symmetric uniform disk model (with $\chi^2_{\nu} = 2.8$).
However, neither the UD nor GD model fully reproduces the trend seen
in the data obtained at high spatial frequencies as can be seen in
Figure~\ref{fig:all_v2}.  Although, the Gaussian model does produce a
better fit, to fully account for the observational signature at the
high spatial frequencies we require a model with an extra degree(s) of
freedom.  A natural extension to a Gaussian model is a two-component
Gaussian model of the form
\begin{equation} \label{eq:dg}
V_{\rm env}^{\rm 2GD}= \rm k_{1} \exp\left[{-\frac{(\pi \theta_{\rm 1} s)^2}{4
      \ln 2}}\right] + \rm (1-k_{1}) \exp\left[{-\frac{(\pi \theta_{\rm 2} s)^2}{4
      \ln 2}}\right], \\
\end{equation}
where $\theta_{\rm 1}$ and $\theta_{\rm 2}$ correspond to the FWHM
values of the two Gaussian components, and $\rm k_{1}$ is the
fractional contribution from the first Gaussian. Similar two-component
Gaussian fits have been applied to other stars by \citet{hof02} and
\citet{pre03}. Combining equations~\ref{eq:model1}, \ref{eq:model2}
and \ref{eq:dg}, and fitting the two-component Gaussian model to the
observations from all three observing seasons results in angular
diameters of $\theta_{\rm 1}$ = 5.64~$\pm$~0.17~mas and $\theta_{\rm
  2}$ = 1.80~$\pm$~0.11~mas, with a fractional contribution from the
stellar photosphere, $\rm c_{p} = 0.70 \pm 0.01$ and $\rm k_{1} = 0.57
\pm 0.02$~(i.e., with the two Gaussian components contributing
approximately equally to the net H$\alpha$ emission).  The best-fit
two-component Gaussian model had the lowest reduced $\chi^2$ value
(yielding $\chi^2_{\nu}$ of 1.5 based on the entire data set) out of
all the models considered in our analysis.  The model curve
corresponding to the two-component Gaussian model is also shown in
Figure~\ref{fig:all_v2}.

The success of the two-component Gaussian model to reproduce the
observational signature from the H$\alpha$ channel is mostly due to
the higher degree of freedom along the radial direction.  In other
words, one can in principle approximate any monotonically decreasing
function as a sum of a large number of Gaussian functions.  In fact,
any other monotonically decreasing function can also be used instead
of a Gaussian.  For example, a sum of a Gaussian distribution and a
uniform disk component of the form
\begin{equation} \label{eq:udgd}
V_{\rm env}^{\rm UD+GD}= l_{1} {2\frac{J_{1}(\pi \theta_{\rm UD}
    s)}{\pi \theta_{\rm UD} s}} + (1-l_{1}) \exp\left[{-\frac{(\pi
      \theta_{\rm GD} s)^2}{4 \ln 2}}\right], \\
\end{equation}
where $\theta_{\rm UD}$ and $\theta_{\rm GD}$ correspond to the
angular diameters of the uniform disk and Gaussian components,
respectively, and $l_{1}$ is the fractional contribution from the
uniform disk, yields equally good fit to the data as the two-component
Gaussian model (resulting in a reduced $\chi^2$ of 1.5).  The best-fit
angular diameters are in this case: $\theta_{\rm
  UD}$~=~3.06~$\pm$~0.15~mas and $\theta_{\rm
  GD}$~=~5.46~$\pm$~0.16~mas, with a fractional contribution from the
stellar photosphere, $c_{\rm p}$, of 0.72 and $l_{1}$~=~0.36.

\subsection{Central Star Diameter} \label{central}

A central star diameter of 0.2~mas was assumed for our model (recall
\S~\ref{inter}). Because varying the adopted diameter for the central
star affects the calculated size of the H$\alpha$-emitting region, we
test the dependency of the angular diameter of the H$\alpha$-emitting
wind structure surrounding P~Cyg on the angular diameter of the
photospheric component.  This is done by fitting
equation~\ref{eq:model1} to the observations with the envelope
component represented by a Gaussian model and the photospheric
component with a UD model for a range of adopted stellar diameters.
Using the observations from the 2007 observing season we find that the
assumed central star diameter of P~Cyg at most has only a 1\% effect
on the derived angular size of the wind structure when the adopted
angular diameter of the central star is varied from 0 to 1~mas.
Therefore, an angular size of the continuum-emitting region that is
underestimated by a factor of five results only in a 1~\% effect on
our best-fit angular diameter of the H$\alpha$-emitting region.

\subsection{Spectroscopic Results}

To test for any possible link between the changes in the strength of
the H$\alpha$ emission line and any variations in the size of the
H$\alpha$-emitting region we examined the spectroscopic data for any
variability on a time-scale comparable to our interferometric
observations.  We found that there was EW variability with an
amplitude $\lesssim$10~\AA~over a time scale of $\sim$2~months, which
is consistent with the short-term variability seen by
\citet{mar01b}. However, the yearly mean H$\alpha$ EW and yearly mean
H$\alpha$ peak signal showed no variation from year to year during our
observing period suggesting relative stability in EW measure on a
year-to-year time scale (see Table \ref{specvartab}).

While our mean spectroscopic data suggest stability on a year-to-year
scale, we cannot rule out variability of H$\alpha$ emission if such
changes are correlated with changes in the continuum level as
suggested by \citet{mar01a}. If the continuum brightened and the
H$\alpha$ has increased in strength by the same fractional amount, it
would appear stable.  Regardless of this apparent mean stability, we
detect high frequency variations on time scales much shorter than one
year.  The upper panel in Figure \ref{fig:EW_vs_V} shows the H$\alpha$
equivalent widths derived from our spectroscopic observations along
with the data published by \citet{mar01a}.

\citet{mar01b} obtained both spectroscopic and photometric
observations of P~Cyg and found that the EW measurements displayed a
slow variability, with an amplitude of about 30~\AA~and a duration of
about 600~days, and a faster variability with an amplitude up to
10~\AA~and duration of 40 to 60~days. While we were able to observe a
variability of up to 10~\AA~in amplitude on a short (month-to-month)
time scale, we did not observe the long-term 30~\AA~amplitude
variability. If the long-term variability discussed in \citet{mar01b}
was present at the time of their observations, it is no longer present
in our data. Therefore, based on our spectroscopic and interferometric
observations, we conclude that P~Cyg is in a quiescent phase of its
lifetime, and it is not going through any drastic changes.

The variability of the H$\alpha$ EW in our data set is similar to the
10~\AA~variation seen in the literature on a time scale of
$\sim$2~months.  Based on our analysis of the equivalent width
measures listed in Table \ref{specvartab}, we conclude that the root
mean square~(RMS) variations are less than 10~\%, and typical closer
to the 2--5~\% range.  Therefore, for the purpose of our study we
assume that the overall emission in the H$\alpha$ line throughout our
interferometric run was stable at the 10~\% level.  We also tested for
variability in the H$\alpha$ strength during a single night and found
it to be virtually non-variable as well (see
Table~\ref{nightspec}). The variations were within the expected
observational uncertainties of $\sim$3~\%, mostly due to uncertainties
associated with continuum level normalization.

\section{Discussion}

\subsection{The Radial Intensity Falloff}

The deviation from a Gaussian shape of the spatially resolved wind
structure of P~Cyg is one of our most intriguing results.  Examining
the squared visibilities in Figure \ref{fig:all_v2}, it is apparent
that the observations at high spatial frequencies cannot be
represented by a single-component Gaussian model.  It is worthwhile to
mention that the Gaussian intensity distribution is still a closer
approximation to the radial intensity distribution from the wind
structure than the uniform disk model (shown with dotted line in
Fig.~\ref{fig:all_v2}).  To our knowledge, this is the first time the
H$\alpha$-emitting region of P~Cyg have been observed to clearly
deviate from a Gaussian shape.

To reproduce our data at the highest spatial frequencies, we use the
two-component Gaussian model as a way of parameterizing the spatially
resolved signature of the H$\alpha$-emitting region of P~Cyg.  This
suggests that the wind structure is indeed more complex in P~Cyg as
compared to Be stars.  Because P~Cyg possess radiatively driven
stellar wind, the velocity structure can be described by the
$\beta$-law of \citet{castor79}, which implies that the wind is
accelerated to its terminal velocity close to the star and then
reaches a constant terminal velocity.  This in turn means that the
density structure of P~Cyg will change at two different rates, one
when the wind is still accelerating and one when the wind has already
reached its terminal velocity at which point the density would be
expected to drop with radial distance as $r^{-2}$.  Although this
cannot be used as a direct explanation for the two component structure
because the H$\alpha$ emission in P~Cyg is optically
thick~\citep{lei88}, by combining the effects of the density
distribution with the radiative transfer effects it should be possible
to use the data presented in this study to directly constrain a
numerical wind model based on a $\beta$ velocity law.  Lastly, if the
two-component wind structure is dominated by optically thick and thin
effects, one could argue that the the inner (optically thick) region
might be better represented by a uniform disk model and the outer
region by a Gaussian model.  This would be in agreement with the
results presented in \S\ref{observation} that showed that the
observations can be equally well represented by a two-component
Uniform-Disk and Gaussian model.

P~Cyg has been spatially resolved in the past using interferometric
techniques. \citet{vak97}, using a single 17.7~m interferometric
baseline, resolved the extended envelope of P~Cyg at the H$\alpha$
emission line. Although their data set was limited and did not allow
them to fit a model more complex than a uniform disk model, the
angular diameter of a uniform disk model they obtained was
$\theta_{\rm UD}$~=~5.52~$\pm$~0.47~mas. They also obtained
complementary H$\alpha$ spectrum on P~Cyg on the same night they
obtained interferometric observations with a H$\alpha$ peak signal of
14.6 $F_{\rm max}/F_{\rm cont}$. This is somewhat weaker than the
$\approx$20 $F_{\rm max}/F_{\rm cont}$ values we obtain based on our
observations from 2005--2007~(recall Fig.~\ref{fig:Pcyg-spectra}),
although the effect of lower spectral resolution in the spectra of
\citet{vak97} cannot be ruled out, which could explain the lower
peak-to-continuum ratio seen in the observation reported by
\citet{vak97}.

Although a uniform disk model is completely inconsistent with our
observations (recall Fig.~\ref{fig:all_v2}), in order to compare our
results to \citet{vak97} at effectively the same limiting spatial
resolution, we fitted a uniform disk model to data from only shortest
baselines (up to 18.9 m).  The best-fit uniform disk angular diameters
ranged from 8.4 to 10.2 mas depending on which season was used in the
fit (see Table~\ref{modelpar}). Therefore, our UD diameter for the
H$\alpha$-emitting wind structure is not consistent with the reported
diameter of 5.52~mas by \citet{vak97}. This discrepancy might be due
to the possibility that \citet{vak97} has modeled the observational
signal that contained both the mostly unresolved central star and
H$\alpha$-emitting envelope with only one uniform disk diameter.  This
would be equivalent to our equation~\ref{eq:model1}, but with the
$c_{\rm p}$ set to 0, which would tend to underestimate the angular
extent of the H$\alpha$ emitting region.  The other possibility is
that the H$\alpha$-emitting region actually grew in size from 1994 to
2007. If we assume that the difference between our values and those
reported by \citet{vak97} to be representative of a real change, then
this results in an expansion rate of the H$\alpha$-emitting
region~(due to changes in opacity) over this thirteen year period of
$\sim$3~km~s$^{-1}$.  This is well within the expected wind velocities
of stars with radiation driven wind structures, especially P~Cyg,
which has a terminal wind velocity almost two magnitudes
larger~\citep{Barlow94,lam85}.  Furthermore, if the H$\alpha$-emitting
region did indeed grow in size, we would expect the H$\alpha$ emission
to be stronger, which is supported by the fact that the $F_{\rm
  max}/F_{\rm c}$ values in Table~\ref{specvartab} that range between
17 and 24 are larger than the $\approx$~15 peak value of the spectrum
taken by \citet{vak97}.

\subsection{Signature of Asymmetry}

\begin{figure}
\plotone{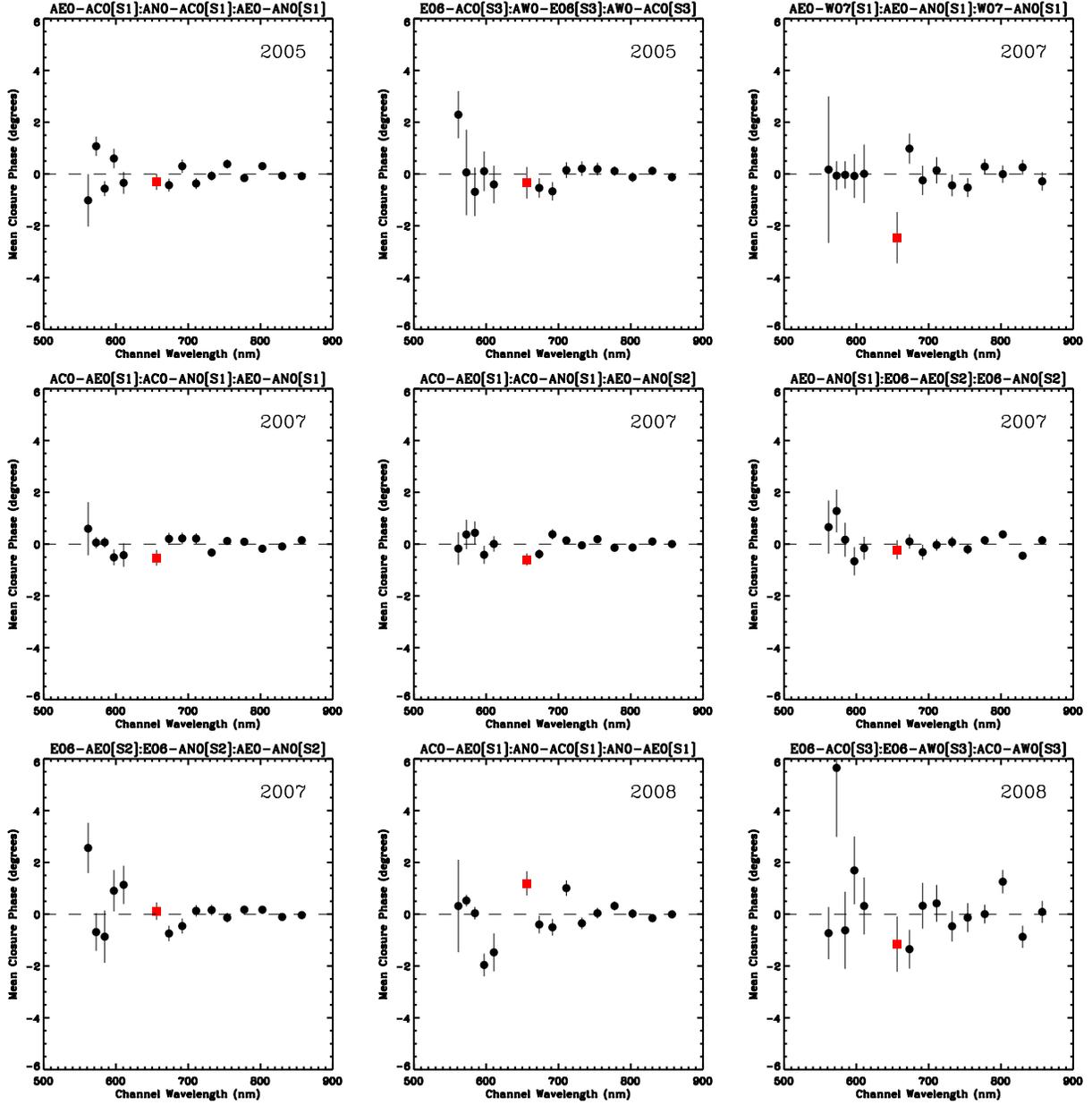}
\caption{Weighted mean closure phases (with a quadratic trend removed
  from the continuum channels) obtained at each possible baseline
  triangle obtained in 2005, 2007 and 2008 observing seasons.  The
  signal from the channel containing the H$\alpha$ emission is marked
  with {\it red squares}.  The closure phases containing the
  AC--AE--AN and E6--AE--AN stations in the 2007 season are plotted in
  two panels since one of the baselines (AE--AN) was observed at two
  independent output beams from the beamcombiner~(indicated in the
  square brackets with a S1 or S2 designator), thereby allowing
  calculation of two closure phase quantities.  The error bars are
  based on the uncertainty of the weighted average.  }
\label{fig:closure_phases}
\end{figure}

To test for the presence of a signature of deviation from
point-symmetry in the signal in the H$\alpha$ channel, we have plotted
the weighted mean closure phases in Figure~\ref{fig:closure_phases}.
If the H$\alpha$ emission originates from an intensity distribution
that is not symmetric across the origin of the photocenter, or
equivalently the H$\alpha$ emitting region is not concentric with the
central star, we would expect to see a non-zero closure phase in the
H$\alpha$ channel.  At a close inspection of
Figure~\ref{fig:closure_phases} we conclude that the H$\alpha$ closure
phases are generally very close to $0^{\circ}$.  Even in the cases
where the mean closure phase deviates from zero values, it is only at
the level of a couple of degrees, which also happens to be only 2 to 3
times the uncertainty of the mean value.  Therefore, we conclude that
based on our observations we do not have a strong signature of
deviation from point symmetry of the H$\alpha$-emitting region.

Our conclusions are not necessarily inconsistent with the a very
similar detection of a subtle phase variation~(at the level of $\sim
30^{\circ}$) across the H$\alpha$ emission line detected by
\citet{vak97} who interpreted that as being produced by a localized
spatial feature~(i.e., a localized blob) within the structure of the
wind.  Assuming that the closure phase variations we detect across the
H$\alpha$ spectral channel in Figure~\ref{fig:closure_phases} are
caused by similar spatial feature (which is not necessarily expected
since the observations were acquired at different epochs), we would
indeed expect to see a much weaker signal across the H$\alpha$ channel
for two reasons.  The first being that we measure the sums of three
phases, which already have the tendency to lower the signal when
negative and positive phases in a triangle are added together.  The
second effect is related to our much wider spectral channel that
contains the entire H$\alpha$ emission line and a significant
contribution from the central star, which most likely can be described
well by a point-symmetric intensity distribution that contains only
real components in the Fourier space (i.e., only real phases).  The
net result is that the real components will tend to lower the net
complex phase detected in the H$\alpha$ channel.

Signatures of clumping in the circumstellar region of P~Cyg on much
larger spatial scales have also been reported.  For example, based on
near-diffraction limited observations using AO system on a 1.52~m
telescope, \citet{che00} reported clumping at the scales of
200--600~mas.  Similarly, P~Cyg was also imaged using the
Multi-Element Radio Linked Interferometer Network (MERLIN) array by
\citet{skin97}.  Their 6~cm observations produced images of the
circumstellar structure on scales of 100--200 mas with a 50 mas
resolution, revealing structural changes on a 40 day timescale along
with flux variations at the 20\% level.  \citet{skin97} argued that
the structural variations could not be attributed to variations in the
mass-loss rate. They suggested recombination within the free-free
emitting region responsible for the radio emission could explain their
observations.  The weak signature of the non-zero closure phase in the
NPOI data, if confirmed, could indicate that the clumpiness observed
by \citet{che00} and \citet{skin97} already originates on scales of
less than 10 mas (less than 25 stellar radii).

\subsection{Photometric Variability}

Lastly, to assess the level of photometric variability during the time
frame covered by our interferometric observations, we used
photoelectric V-band observations from the American Association of
Variable Star Observers.  The $V$-band light curve from 2005 to 2009
is shown in Figure~\ref{fig:EW_vs_V} where we only include data for
which the measured check star~(HD~193369) magnitude was no more than
0.04~mag away from the mean value of 5.573~mag.  We see both a mean
seasonal-timescale variability on the order of $\sim 0.02$ mag, and
very short term variability (for observations taken within days to
weeks) at the level of $\pm 0.05$~mag.  Although the short-term
variability appears to have large scatter, the photometric quality of
the AAVSO PEP data is generally good, with internal errors on the
order of 5--10 milli-magnitudes.  Therefore, based on the photometric
data we conclude that although the mean photometric level appears to
change by less than 0.02~mag, short term variations upwards of 0.1~mag
cannot be ruled out.  This implies that the continuum level in P~Cyg
could vary up to 10~\% level, which would affect directly the EW
measures at the same level.  This also strongly suggests that the
variations in EW measures seen in Table~\ref{specvartab} are caused by
variations in the continuum level and not the emission component
itself.

\section{Summary}

In general, mass-loss rates and stellar wind properties are extremely
important characteristics of LBVs. Mass-loss rates for P~Cyg have
already been estimated through several methods, but none of them rely
on spatially resolved H$\alpha$-emitting inner region.  In this
observational study we have presented the most comprehensive
interferometric data set on P~Cyg to date.  Our results provide a new
and independent source of observational constraints that can be
utilized when determining the properties of the wind structure in the
context of the predicted H$\alpha$ intensity distribution on the sky.

We have fitted circularly symmetric uniform disk, Gaussian, and
two-component models to the interferometric observations and have
shown how the H$\alpha$-emitting wind structure of P~Cyg cannot be
represented by a simple one component model.  We conclude that the
wind structure can be represented fully by a two-component Gaussian
model with angular diameters of $\theta_{1}$~=~5.64~$\pm$~0.21~mas and
$\theta_{2}$~=~1.80~$\pm$~0.13~mas.  The interferometric signature can
also be equally well represented with a two-component model consisting
of a uniform disk and a Gaussian intensity distribution.  In that case
the angular diameters are $\theta_{\rm UD}$~=~3.06~$\pm$~0.15~mas and
$\theta_{\rm GD}$~=~5.46~$\pm$~0.16~mas.  While we cannot conclude if
a double Gaussian model or a uniform-disk plus a Gaussian model better
represents the wind structure of P~Cyg, it is clear that both models
represent the wind structure better than a single Gaussian or uniform
disk model.  We conclude that P~Cyg's wind structure is complex,
possibly containing multiple layers of varying opacities.  The data
presented in this study might be used to directly constrain wind
models of P~Cyg, perhaps allowing for an independent observational
constraint on the $\beta$-law.

Based on our spectroscopic and interferometric observations we
conclude that P~Cyg is stable at the 10~\% level.  Based on the
photometric observations we also rule out the possibility that large
changes in the H$\alpha$ EW measure were masked by changes in the
continuum level.  The lack of variability at a significant level
combined with the spatially resolved H$\alpha$-emitting region that
appears to be stable between the observing seasons is suggestive of an
extra-quiescent phase of P~Cyg between 2005 and 2008.

\acknowledgements
\small

The Navy Prototype Optical Interferometer is a joint project of the
Naval Research Laboratory and the U.S. Naval Observatory, in
cooperation with Lowell Observatory, and is funded by the Office of
Naval Research and the Oceanographer of the Navy.  We thank Anatoly
Miroshnichenko and Glen Williams for valuable suggestions on how to
improve the modeling section of this work.  We are very grateful for
the very detailed and constructive comments and suggestions we have
received from the anonymous referee.  We would also like to thank the
Lowell Observatory for the generous telescope time allocation on the
John S. Hall Telescope and Nevena Markova for providing us with
electronic files of the published data.  We acknowledge with thanks
the variable star observations of the AAVSO Photoelectric Photometry
program contributed by observers worldwide and used in this research.
A.B. and C.T. would like to thank Central Michigan University for the
financial support for this project.  C.T. acknowledges, with thanks,
grant support from Research Corporation for Science Advancement.  This
research has made use of NASA's Astrophysics Data System Bibliographic
Services, and the SIMBAD database, operated at CDS, Strasbourg,
France.

{\it Facilities:} \facility{NPOI}, \facility{Hall (Solar Stellar
  Spectrograph)}

\clearpage



\clearpage

\clearpage

\clearpage

\clearpage

\clearpage

\clearpage

\clearpage

\begin{table}
\begin{center}
\parbox{2.9in}{\caption{NPOI Observing Log of P~Cyg\label{log}}}
\begin{tabular}{lc}
\hline\hline
UT Date \hspace{3.5cm} & \# of H$\alpha$ squared\\ & visibilities\\\hline
2005 Jun 2               \dotfill &       48\\
2005 Jun 3               \dotfill &       19\\
2005 Jun 4               \dotfill &       48\\
2005 Jun 6               \dotfill &       18\\
2005 Jun 8               \dotfill &       20\\
2005 Jun 9               \dotfill &       84\\
2005 Jun 11              \dotfill &       19\\
2005 Jun 13              \dotfill &       50\\ 
2005 Jun 16              \dotfill &       90\\\hline
2007 Jul 1               \dotfill &       20\\
2007 Jul 2               \dotfill &       20\\
2007 Jul 4               \dotfill &       20\\
2007 Jul 12              \dotfill &       15\\
2007 Aug 8               \dotfill &       24\\
2007 Aug 9               \dotfill &       16\\
2007 Aug 10              \dotfill &       66\\
2007 Aug 11              \dotfill &       56\\
2007 Aug 12              \dotfill &       28\\
2007 Aug 18              \dotfill &       72\\
2007 Aug 19              \dotfill &       31\\
2007 Aug 20              \dotfill &       82\\
2007 Aug 21              \dotfill &       33\\
2007 Aug 22              \dotfill &       92\\
2007 Aug 23              \dotfill &       17\\
2007 Aug 24              \dotfill &       12\\
2007 Aug 25              \dotfill &       41\\
2007 Aug 27              \dotfill &       12\\
2007 Aug 29              \dotfill &       12\\
2007 Sep 5               \dotfill &       36\\\hline
2008 Jun 27              \dotfill &       16\\
2008 Jun 28              \dotfill &       40\\
2008 Jun 30              \dotfill &       36\\
2008 Jul 2               \dotfill &       28\\
2008 Jul 3               \dotfill &       60\\
2008 Jul 7               \dotfill &       64\\
2008 Jul 8               \dotfill &       89\\
2008 Jul 9               \dotfill &      100\\\hline
Total                    \dotfill &     1534\\\hline
\end{tabular}
\end{center}
\end{table}

\clearpage

\begin{table}[htp]
\caption[]{Calibrated H$\alpha$ Squared Visibilities of P~Cyg}
\label{tab:v2data}
\begin{tabular}{crrcc} \hline\hline
 Julian Date        & Spatial Frequency $u$ & Spatial Frequency $v$ &               &         \\
(JD $-$ 2,450,000) & ($10^6$ cycles/radian)  & ($10^6$ cycles/radian)  &   $V_{\rm{H}\alpha}^2$ & Baseline \
\\ \hline
3523.863  &  $ -18.364$  &  $  19.654$  &  0.666 $\pm$ 0.028 & AC-AE \\
3523.863  &  $ -11.184$  &  $ -32.672$  &  0.623 $\pm$ 0.040 & AC-AN \\
3523.863  &  $  -7.180$  &  $  52.326$  &  0.615 $\pm$ 0.017 & AE-AN \\
3523.863  &  $ -37.854$  &  $  29.298$  &  0.567 $\pm$ 0.023 & AC-E6 \\
3523.863  &  $  67.722$  &  $ -25.810$  &  0.512 $\pm$ 0.026 & AW-E6 \\
3523.863  &  $  29.868$  &  $   3.488$  &  0.671 $\pm$ 0.045 & AC-AW \\
3523.891  &  $ -21.636$  &  $  17.482$  &  0.705 $\pm$ 0.023 & AC-AE \\
3523.891  &  $  -8.106$  &  $ -33.720$  &  0.633 $\pm$ 0.039 & AC-AN \\
3523.891  &  $ -13.530$  &  $  51.202$  &  0.515 $\pm$ 0.023 & AE-AN \\
3523.891  &  $ -43.328$  &  $  24.890$  &  0.526 $\pm$ 0.025 & AC-E6 \\
3523.891  &  $  74.493$  &  $ -18.088$  &  0.501 $\pm$ 0.033 & AW-E6 \\
\hline
\end{tabular}\\[0.9ex]
\parbox{6.0in}{\footnotesize \quad {\sc Note.} ---
Table~\ref{tab:v2data} is published in its entirety in the electronic
edition of the Astronomical Journal.}
\end{table}

\clearpage

\begin{table}
\begin{center}
\parbox{3.2in}{\caption{Spectroscopic Variability for P~Cyg\label{specvartab}}}
\begin{tabular}{lcccc}
\hline\hline
UT Date & --EW (\AA) & \% Change$^{\dag}$ & $\frac{F_{max}}{F_{c}}$ & \% Change$^{\dag}$ \\
\hline 
2005 Apr 1  & 82.1           & 3.5 & 20.8           & 6.7 \\
2005 May 19 & 74.9           & 5.7 & 18.7           & 4.0 \\
2005 Jun 28 & 81.0           & 2.0 & 20.8           & 6.7 \\
2005 Jul 16 & 80.3           & 1.1 & 20.8           & 6.9 \\
2005 Aug 22 & 79.7           & 0.5 & 19.7           & 1.1 \\
2005 Sep 16 & 78.8           & 0.8 & 19.1           & 2.0 \\
2005 Sep 17 & 81.3           & 2.4 & 19.6           & 0.6 \\
2005 Sep 27 & 81.0           & 2.1 & 19.5           & 0.1 \\
2005 Oct 11 & 81.0           & 2.1 & 18.9           & 3.1 \\
2005 Nov 8  & 73.6           & 7.3 & 17.0           & 12.9\\ \hline
2005 Mean   & 79.4~$\pm$~0.9 &     & 19.5~$\pm$~0.4 &     \\ \hline

2006 Jun 10 & 81.4           & 7.3 & 19.7           & 7.3 \\
2006 Aug 3  & 70.1           & 7.7 & 16.6           & 9.4 \\
2006 Sep 6  & 75.3           & 0.8 & 17.9           & 2.7 \\
2006 Oct 3  & 70.3           & 7.3 & 17.2           & 6.2 \\
2006 Oct 28 & 78.1           & 2.9 & 19.0           & 3.3 \\
2006 Oct 31 & 77.4           & 2.0 & 19.0           & 3.4 \\
2006 Nov 1  & 78.6           & 3.5 & 19.2 	    & 4.4 \\ \hline
2006 Mean   & 75.9~$\pm$~1.6 &     & 18.4~$\pm$~0.4 &     \\ \hline

2007 May 3  & 70.5           & 10.8 & 17.3          & 17.7 \\
2007 Jun 4  & 75.8           & 4.1  & 18.7          & 11.1 \\
2007 Jun 22 & 80.0           & 1.2  & 19.7          & 6.4  \\
2007 Oct 23 & 83.4           & 5.5  & 22.5          & 6.8  \\
2007 Oct 24 & 78.0           & 1.3  & 21.1          & 0.3  \\
2007 Oct 27 & 76.5           & 3.2  & 20.9          & 0.6  \\
2007 Oct 28 & 74.7           & 5.6  & 20.2          & 3.9  \\
2007 Dec 18 & 80.9           & 2.4  & 22.6          & 7.5  \\
2007 Dec 19 & 85.5           & 8.2  & 23.7          & 12.7 \\
2007 Dec 20 & 85.2           & 7.8  & 23.7          & 12.5 \\ \hline
2007 Mean   & 79.0~$\pm$~1.6 &      & 21.0~$\pm$~0.7 &     \\ \hline
\end{tabular}
\parbox{4.5in}{\footnotesize $^{\dag}$ Changes calculated with respect to yearly mean EW and peak strength.}
\end{center}
\end{table}

\clearpage

\begin{table}[htp]
\caption[]{Best-Fit Model Parameters for P~Cyg}
\label{modelpar}
\small
\begin{tabular}{clccccr} \hline\hline
Model \hspace{2.1cm} & Season & Diameter (mas) & 2$^{\rm nd}$ Diameter (mas) & $k_1$ or $l_1$ & ${c}_{\rm p}$ & ${\chi}^2_\nu$\\
\hline
Uniform Disk$^{\dag}$  \dotfill & 2005  &     9.41 $\pm$ 2.50 &  \ldots  &  \ldots   &      0.84 $\pm$ 0.03 &  \\
Uniform Disk           \dotfill & 2005  &     6.74 $\pm$ 0.05 &  \ldots  &  \ldots   & 0.78 $\pm$ 0.01 & 2.6\\
Gaussian               \dotfill & 2005  &     3.97 $\pm$ 0.05 &  \ldots  &  \ldots   & 0.75 $\pm$ 0.01 & 1.5\\
Double-Gaussian        \dotfill & 2005  &     8.7  $\pm$ 5.3  & 2.08 $\pm$ 0.15 & 0.41 $\pm$ 0.03 & 0.70 $\pm$ 0.01 & 1.2\\
\hline
Uniform Disk$^{\dag}$  \dotfill & 2007  &    10.17 $\pm$ 0.32 &  \ldots  &  \ldots   & 0.84 $\pm$ 0.01 &  \\
Uniform Disk           \dotfill & 2007  &     7.37 $\pm$ 0.03 &  \ldots  &  \ldots   & 0.79 $\pm$ 0.01 & 3.7\\
Gaussian               \dotfill & 2007  &     4.37 $\pm$ 0.03 &  \ldots  &  \ldots   & 0.76 $\pm$ 0.01 & 2.4\\
Double-Gaussian        \dotfill & 2007  &     5.33 $\pm$ 0.15 & 1.53 $\pm$ 0.15 & 0.60 $\pm$ 0.02 & 0.70 $\pm$ 0.01 & 1.7\\
\hline
Uniform Disk$^{\dag}$  \dotfill & 2008  &     8.40 $\pm$ 1.50 &  \ldots  &  \ldots   & 0.79 $\pm$ 0.04 &  \\
Uniform Disk           \dotfill & 2008  &     7.06 $\pm$ 0.06 &  \ldots  &  \ldots   & 0.78 $\pm$ 0.01 & 1.5\\
Gaussian               \dotfill & 2008  &     4.37 $\pm$ 0.07 &  \ldots  &  \ldots   & 0.75 $\pm$ 0.01 & 1.3\\
Double-Gaussian$^{\ddag}$ \dotfill & 2008  &      10  $\pm$ 9    & 2.17 $\pm$ 0.07 & 0.50 $\pm$ 0.02 & 0.70 & 1.1\\
\hline
Uniform Disk$^{\dag}$  \dotfill & All   &     9.83 $\pm$ 0.26 &  \ldots  &  \ldots   & 0.84 $\pm$ 0.01 &  \\
Uniform Disk           \dotfill & All   &     7.21 $\pm$ 0.03 &  \ldots  &  \ldots   & 0.79 $\pm$ 0.01 & 2.8\\
Gaussian               \dotfill & All   &     4.29 $\pm$ 0.03 &  \ldots  &  \ldots   & 0.76 $\pm$ 0.01 & 1.9\\
Double-Gaussian        \dotfill & All   &     5.64 $\pm$ 0.17 & 1.80 $\pm$ 0.11 & 0.57 $\pm$ 0.02 & 0.70 $\pm$ 0.01 & 1.5\\
Uniform Disk + Gaussian\dotfill & All   &     3.06 $\pm$ 0.15 & 5.46 $\pm$ 0.16 & 0.36 $\pm$ 0.02 & 0.72 $\pm$ 0.01 & 1.5\\
\hline
\end{tabular}\\
\parbox{7.0in}{\footnotesize $^{\dag}$ Models fitted to data for
  baselines up to 18.9~m. \\ $^{\ddag}$ There were insufficient data at
  the longest baselines to fit for four model parameters and instead a
  fit was obtained with $c_{\rm p}$ fixed at 0.70.}
\end{table}

\clearpage

\begin{table}
\begin{center}
\parbox{4.0in}{\caption{Single Night Spectroscopic Variability for P~Cyg\label{nightspec}}}
\begin{tabular}{lcccc}
\hline \hline
UT Date & --EW (\AA) & \% Change$^{\dag}$ & $\frac{F_{max}}{F_{c}}$ & \% Change$^{\dag}$ \\
\hline 
2007 Dec 20 & 85.2 & 0.1 & 23.7 & 0.0 \\
2007 Dec 20 & 85.9 & 0.9 & 23.8 & 0.7 \\
2007 Dec 20 & 84.4 & 0.9 & 23.5 & 0.7 \\
2007 Dec 20 & 85.6 & 0.5 & 23.8 & 0.4 \\
2007 Dec 20 & 85.2 & 0.1 & 23.7 & 0.0 \\
2007 Dec 20 & 85.8 & 0.7 & 23.8 & 0.6 \\
2007 Dec 20 & 86.0 & 1.0 & 23.8 & 0.6 \\
2007 Dec 20 & 83.4 & 2.1 & 23.4 & 1.2 \\
2007 Dec 20 & 85.0 & 0.2 & 23.6 & 0.2 \\
2007 Dec 20 & 85.2 & 0.0 & 23.7 & 0.0 \\
2007 Dec 20 & 83.4 & 2.1 & 23.4 & 1.2 \\
2007 Dec 20 & 84.7 & 0.6 & 23.6 & 0.4 \\
2007 Dec 20 & 85.3 & 0.1 & 23.6 & 0.1 \\
2007 Dec 20 & 85.5 & 0.4 & 23.7 & 0.3 \\
2007 Dec 20 & 85.4 & 0.2 & 23.7 & 0.1 \\
2007 Dec 20 & 85.7 & 0.6 & 23.8 & 0.5 \\
2007 Dec 20 & 86.1 & 1.1 & 23.8 & 0.6 \\
2007 Dec 20 & 85.5 & 0.4 & 23.7 & 0.3 \\
2007 Dec 20 & 85.1 & 0.1 & 23.6 & 0.3 \\ \hline
2007 Dec 20 Mean & 85.2~$\pm$~0.2 &  & 23.67~$\pm$~0.03 &  \\
\hline
\end{tabular}
\parbox{5.0in}{\footnotesize $^{\dag}$ Changes calculated with respect to nightly mean. All spectra were obtained with 180~s of integration.}
\end{center}
\end{table}

\end{document}